\documentstyle[12pt]{article}
\begin{document}
\title{LAW WITHOUT LAW}
\author{B.G. Sidharth\\
International Institute for Applicable Mathematics \& Information Sciences\\
Hyderabad (India) \& Udine (Italy)\\
B.M. Birla Science Centre, Adarsh Nagar, Hyderabad - 500 063
(India)}
\date{}
\maketitle
\begin{abstract}
We consider a model for spacetime in which there is an ubiquitous
background Dark Energy which is the Zero Point Field. This is
further modeled in terms of a Weiner process that leads to a Random
or Brownian characterization. Nevertheless we are able to recover
meaningful physics, very much in the spirit of Wheeler's Law without
Law, that is laws emerging from an underpinning of lawlessness.
\end{abstract}
\section{Introduction}
From the beginning of modern science, the universe has been
considered to be governed by rigid laws which therefore, in a sense,
made the universe somehow deterministic. However, it would be more
natural to expect that the underpinning for these laws would be
random,unpredictable and spontaneous rather than enforced events.
This alternative but historical school of thought is in the spirit
of Prigogine's,
"Order out of chaos"\cite{r1}.\\
Prigogine notes, "As we have already stated, we subscribe to the
view that classical science has now reached its limit. One aspect of
this transformation is the discovery of the limitations of classical
concepts that imply that a knowledge of the world "as it is" was
possible. The omniscient beings, Laplace's or Maxwell's  demon, or
Einstein's God, beings that play such an important role in
scientific reasoning, embody the kinds of extrapolation physicists
thought they were allowed to make. As randomness, complexity, and
irreversibility enter into physics as objects of positive knowledge,
we are moving away from this rather naive assumption of a direct
connection between our description of the world and the world
itself. Objectivity in theoretical physics takes on a more subtle
meaning. ...Still there is only one type of change surviving in
dynamics, one "process", and that is motion... It is interesting to
compare dynamic change with the atomists' conception of change,
which enjoyed considerable favor at the time Newton formulated his
laws. Actually, it seems that not only Descartes, Gessendi, and
d'Alembert, but even Newton himself believed that collisions between
hard atoms were the ultimate, and perhaps the only, sources of
changes of motion. Nevertheless, the dynamic and the atomic
descriptions differ radically. Indeed, the continuous nature of the
acceleration described by the dynamic equations is in sharp contrast
with the discontinuous, instantaneous collisions between hard
particles. Newton had already noticed that, in contradiction to
dynamics, an irreversible loss of motion is involved in each hard
collision. The only reversible collision--that is, the only one in
agreement with the laws of dynamics--is the "elastic,"
momentum-conserving collision. But how can the complex property of
"elasticity" be applied to atoms that are supposed to be the
fundamental elements of nature?\\
"On the other hand, at a less technical level, the laws of dynamic
motion seem to contradict the randomness generally attributed to
collisions between atoms. The ancient philosophers had already
pointed out that any natural process can be interpreted in many
different ways in terms of the motion of and collisions between
atoms."\\
 In the words of Wheeler\cite{r2},
we seek ultimately a "Law without Law."  Laws are an apriori blue
print within the constraints of which, the universe evolves. The
point can be understood
in the words of Prigogine \cite{prig}\\
"...This problem is a continuation of the famous controversy between
Parmenides and Heraclitus. Parmenides insisted that there is nothing
new, that everything was there and will be ever there. This
statement is paradoxical because the situation changed before and
after he wrote his famous poem. On the other hand, Heraclitus
insisted on change. In a sense, after Newton's dynamics, it seemed
that Parmenides was right, because Newton's theory is a
deterministic theory and time is reversible. Therefore nothing new
can appear. On the other hand, philosophers were divided. Many great
philosophers shared the views of Parmenides. But since the
nineteenth century, since Hegel, Bergson, Heidegger, philosophy took
a different point of view. Time is our existential dimension. As you
know, we have inherited from the nineteenth century two different
world of views. The world view of dynamics, mechanics and the world
view of thermodynaics."\\
It may be mentioned that subsequent developments in Quantum Theory,
including Quantum Field Theory are in the spirit of the former.
Einstein himself believed in this view of what may be called
deterministic time - time that is also reversible. On the other hand
Heraclitus's point of view was in the latter spirit. His famous
dictum was, "You never step into the same river twice", a point of
view which was endorsed by earlier ancient Indian thought. This has
been the age old tussle between "being" and "becoming".\\
As Wheeler put it, (loc.cit), "All of physics in my view, will be
seen someday to follow the pattern of thermodynamics and statistical
mechanics, of regularity based on chaos, of "law without law".
Specifically, I believe that everything is built higgledy-piggledy
on the unpredictable outcomes of billions upon billions of
elementary quantum phenomena, and that the laws and initial
conditions of physics arise out of this chaos by the action of a
regulating principle, the
discovery and proper formulation of which is the number one task...."\\
The reason this approach is more natural is, that otherwise we would
be lead to ask, "from where have these laws come?" unless we either
postulate a priori laws or we take shelter behind an anthropic
argument. An interesting but neglected body of work in the past few
decades is that of Random or Stochastic Mechanics and
Electrodynamics. It may be mentioned that a considerable amount of
work has been done in this direction by Nelson, Landau, Prugovecki,
the author and others\cite{r3}-\cite{r29}, who have tried to derive
the Schrodinger equation, the Klein-Gordon equation and even the
Dirac equation from stochastic considerations, and in general
develop an underpinning of stochastic mechanics and stochastic
electrodynamics. The literature is vast and some of the references
given cite an extensive bibliography. A few of these approaches have
been very briefly touched upon in Cf.ref.\cite{uof}. However, all
these derivations contain certain assumptions whose meaning has been
unclear. We will see examples of this in the sequel. In any case, we
will argue that the seeds of a new world view,
of the paradigm shift are to be found here in these considerations.\\
In the above context, we propose below that purely stochastic
processes lead to minimum space-time intervals of the order of the
Compton wavelength and time, whose considerable significance will be
seen and it is this circumstance that underlies quantum phenomena
and cosmology, and, in the thermodynamic limit in which $N$, the
number of particles in the universe $\to \infty$, classical
phenomena and Quantum Theory as well. In the process, we will obtain
a rationale for some of the ad hoc assumptions referred to above.\\
In the older, and more popular world view, spacetime has generally
been taken to be a differentiable manifold with an Euclidean
(Galilean) or Minkowskian or Riemannian character. Though the
Heisenberg Principle in Quantum Theory forbids arbitrarily small
space time intervals, the above continuum character with space time
points has been taken for granted even in Quantum Field Theory. In
fact if we accept the proposition that what we know of the universe
is a result of our measurement (which includes our perception), and
that measurements are based on quantifiable units, then it becomes
apparent that a continuum is at best an idealization. This was the
reason behind the paradox of the point electron which was
encountered in the Classical theory of the electron, as we saw. It
was also encountered as is well known in Dirac's Quantum Mechanical
treatment of the relativistic, spinning electron in which the
electron showed up with the velocity of
light.\\
Quantum Mechanics has lived with this self contradiction\cite{r30}.
In this schizophrenic existence, the wave function follows a
deterministic (time reversible) equation, while the result of a
measurement, without which no information is retrievable, follows
from an acausal "collapse of the wave function" yielding one of the
many permissible eigen values, in an unpredictable but probabilistic
manner. Indeed it has been suggested by Snyder, Lee and others that
the infinities which plague Quantum Field Theory are symptomatic of
the fact that space time has a granular or discrete rather than
continuous character. This has lead to a consideration of extended
particles\cite{r31}-\cite{r37} \cite{r38}, as against point
particles of conventional theory. Wheeler's space time foam and
strings\cite{r39}-\cite{r43} are in this class, with a minimum cut
off at the Planck scale. As 't Hooft notes, \cite{hooft} "It is
somewhat puzzling to the present author why the lattice structure of
space and time had escaped attention from other investigators up
till now..." We will return to this point later.\\
All this has also lead to a review of the conventional concept of a
rigid background space time. More recently \cite{r44}-\cite{r46}, it
has been pointed out by the author that it is possible to give a
stochastic underpinning to space time and physical laws. This is in
the spirit of Wheeler's, "Law without Law" \cite{r2} alluded to. In
fact in a private communication to the author, Prof. Prigogine
wrote, "...I agree with you that spacetime has a stochastic
underpinning".
\section{The Emergence of Space-Time}
We will later briefly survey some models for spacetime. For the
moment our starting point is the well known fact that in a random
walk, the average distance $l$ covered at a stretch is given by
\cite{r47}
\begin{equation}
l = R/\sqrt{N}\label{e1}
\end{equation}
where $R$ is the dimension of the system and $N$ is the total number
of steps. We get the same relation in Wheeler's famous travelling
salesman problem and  similar problems\cite{r48} The interesting
fact that equation (\ref{e1}) is true in the universe itself with
$R$ the radius of the universe $\sim 10^{28}cm,N$ the number of the
elementary particles in the universe $\sim 10^{80}$ and $l$ the
Compton wavelength of the typical elementary particle, for example
the pion $\sim 10^{-13}cm$ had been noticed a long time
ago\cite{r49}. From a different point of view, it is one of the
cosmic "coincidences" or Large Number relations, pointed out by
Weyl, Eddington and others. In this context, equation (\ref{e1})
which has been generally considered to be accidental (along with
other such relations which we will encounter), will be shown to
arise quite naturally in a cosmological scheme based on
fluctuations. We would like to stress that we encounter the Compton
wavelength as an important and fundamental minimum unit of length
and will return
recurrently to this theme.\\
It may be mentioned that a minimum time interval, the chronon, has
been considered earlier in a different context by several authors as
we will see very soon. What distinguishes Quantum Theory from
Classical Physics is as pointed out, the role of the resolution of
the observer or observing apparatus. What appears smooth at one
level of perception, may turn out to be very irregular on a closer
examination. Indeed as noted by Abbot and Wise\cite{r50}, in this
respect the situation is similar to everywhere continuous but non
differentiable curves, the fractals of Mandelbrot \cite{r51}. This
again is tied up with the Random Walk or Brownian character of the
Quantum path as noted by Sornette and others\cite{r52}-\cite{r59}:
At scales larger than the Compton wavelength but smaller than the de
Broglie wavelength, the Quantum paths have the fractal dimension 2
of Brownian paths (cf. also Nottale,\cite{r60}). This will be
touched upon briefly in Section 6.\\
This irregular nature of the
Quantum Mechanical path was noticed by Feynman \cite{fh} "...these
irregularities are such that the 'average' square velocity does not
exist, where we have used the classical analogue
in referring to an 'average'.\\
"If some average velocity is defined for a short time interval
$\Delta t$, as, for example, $|x(t + \Delta t) - x(t)|/\Delta t$,
the "mean" square value of this is $- \hbar / (m\Delta t)$. That is,
the "mean" square value of a velocity averaged over a short time
interval is finite, but its value becomes larger as the
interval becomes shorter.\\
It appears that quantum-mechanical paths are very irregular.
However, these irregularities average out over a reasonable length
of time to produce a reasonable drift, or "average" velocity,
although for short intervals of time the "average" value of the
velocity is very high..."\\
This as we will see was Dirac's conclusion too, and indeed his
explanation for the luminal velocity of the point electron and the
non
Hermiticity of its position operator in his relativistic electron theory.\\
Two important characteristics of the Compton wavelength have to be
re-emphasized (Cf.\cite{r46}): On the one hand with a minimum space
time cut off at the Compton wavelength, as we will see, we can
recover by a simple coordinate shift the Dirac structure for the
equation of the electron, including the spin half. In this sense the
spin half, which is purely Quantum Mechanical will be seen to be
symptomatic of the minimum space time cut off, as is also suggested
by the zitterbewegung interpretation of Dirac (in terms of the
Uncertainty Principle), Hestenes and others (Cf. discussion in
\cite{uof}). The zitterbewegung is symptomatic of the fact that by
the Heisenberg Uncertainty Principle, physics begins only after an
averaging over the minimum space time intervals. This is also
suggested by stochastic models of Quantum Mechanics referred to,
both non relativistic and relativistic as also Feynman's Path
Integral formulation. We will
comment upon in the sequel.\\
On the other hand, we will see that (\ref{e1}) and a similar
equation for the Compton time in terms of the age of the universe,
viz.,
\begin{equation}
T \approx \sqrt{N} \tau\label{e2}
\end{equation}
can be the starting point for a unified scheme for physical
interactions and indeed a cosmology that is not only consistent with
observation in which we will deduce the Large Number coincidences
referred to, but also predicted in 1997 an accelerating expanding
universe when the ruling paradigm was exactly the opposite. We will
see this in the next Chapter in detail. The Large Number relations
also include a mysterious formula \cite{r61}, connecting the pion
mass and the Hubble constant which we will deduce. It has to be
pointed out \cite{r48} that in the spirit of Wheeler's travelling
salesman's "practical man's minimum" length that the Compton scale
plays such a role, and that space time is like Richardson's
delineation of a jagged coastline \cite{r51} with a thick brush, the
thickness of the brush being
comparable to the Compton scale.\\
What Richardson found was that the length of the common land
boundaries claimed by Portugal and Spain as also Netherlands and
Belgium, differed by as much as 20$\%$! The answer to this
non-existent border dispute lies in the fact that we are carrying
over our concepts of smooth curves or rectifiable arcs to the
measurement of real life jagged boundaries or coastlines. As far as
these latter are concerned, as Mandelbrot puts it \cite{r51} "The
result is most peculiar; coastline length turns out to be an elusive
notion that slips between the fingers of one who wants to grasp it.
All measurement methods ultimately lead to the conclusion that the
typical coastline's length is very large and so ill determined that
it is best considered infinite....." This is where Hansdorf
dimension or the fractal dimension referred to earlier comes in-- we
are approximating a higher
dimensional curve by a one dimensional curve.\\
Space time, rather than being a smooth continuum, is more like a
fractal Brownian curve, what may be called Quantized Fractal
Spacetime. All this has been recognized by some scholars, at least
in spirit. As V.L. Ginzburg puts it \cite{gins} "The special and
general relativity theory, non-relativistic quantum mechanics and
present theory of quantum fields use the concept of continuous,
essentially classical, space and time (a point of spacetime is
described by four coordinates $x_l = x,y,z, ct$ which may vary
continuosly). But is this concept valid always? How can we be sure
that on a "small scale" time and space do not become quite
different, somehow fragmentized, discrete, quantized? This is by no
means a novel question, the first to ask it was, apparently Riemann
back in 1854 and it has repeatedly been discussed since that time.
For instance, Einstein said in his well known lecture "Geometry and
Experience" in 1921: 'It is true that this proposed physical
interpretation of geometry breaks down when applied immediately to
spaces of submolecular order of magnitude. But nevertheless, even in
questions as to the constitution of elementary particles, it retains
part of its significance. For even when it is a question of
describing the electrical elementary particles constituting matter,
the attempt may still be made to ascribe physical meaning to those
field concepts which have been physically defined for the purpose of
describing the geometrical behavior of bodies which are large as
compared with the molecule. Success alone can decide as to the
justification of such an attempt, which postulates physical reality
for the fundamental principles of Riemann's geometry outside of the
domain of their physical definitions. It might possibly turn out
that this extrapolation has no better warrant than the extrapolation
of the concept of
temperature to parts of a body of molecular order of magnitude'.\\
"This lucidly formulated question about the limits of applicability
of the Riemannian geometry (that is, in fact macroscopic, or
classical, geometric concepts) has not yet been answered. As we move
to the field of increasingly high energies and, hence to "closer"
collisions between various particles the scale of unexplored space
regions becomes smaller. Now we may possibly state that the usual
space relationships down to the distance of the order of
$10^{-15}cm$ are valid, or more exactly, that their application does
not lead to inconsistencies. It cannot be ruled out that, the limit
is nonexistent but it is much more likely that there exists a
fundamental (elementary) length $l_0 \leq 10^{-16} - 10^{-17}cm$
which restricts the possibilities of classical, spatial description.
Moreover, it seems reasonable to assume that the fundamental length
$l_0$ is, at least, not less than the gravitational length $l_g =
\sqrt{Gh/c^3} \sim 10^{-33}cm$.\\
"... It is probable that the fundamental length would be a "cut-off"
factor which is essential to the current quantum theory: a theory
using a fundamental length automatically excludes divergent
results".\\
Einstein himself was aware of this possibility. As he observed
\cite{einstein}, "... It has been pointed out that the introduction
of a spacetime continuum may be considered as contrary to nature in
view of the molecular structure of everything which happens on a
small scale. It is maintained that perhaps the success of the
Heisenberg method points to a purely algebraic method of description
of nature that is to the elimination of continuous functions from
physics. Then however, we must also give up, by principle the
spacetime continuum. It is not unimaginable that human ingenuity
will some day find methods which will make it possible to proceed
along such a path. At present however, such a program looks like an
attempt to breathe in
empty space".\\
To analyse this further, we observe that space time given by $R$ and
$T$ of (\ref{e1}) and (\ref{e2}) represents a measure of dispersion
in a normal distribution: Indeed if we have a large collection of
$N$ events (or steps) of length $l$ or $\tau$, forming a normal
distribution, then the dispersion $\sigma$ is given by precisely the
relation (\ref{e1}) or
(\ref{e2}).\\
The significance of this is brought out by the fact that the
universe is a collection of $N$ elementary particles, infact
typically pions of size $l$, as seen above. We consider space time
not as an apriori container of these particles but rather as a
Gaussian collection of these particles, a Random
Heap. At this stage, we do not even need the concept of a continuum.\\
In this scheme the probability distribution has a width or
dispersion $\sim \frac{1}{\sqrt{N}}$ (Cf. ref.\cite{r65,r66,r67}),
that is the fluctuation (or dispersion) in the number of particles
$\sim \sqrt{N}$.
This immediately leads to equations (\ref{e1}) and (\ref{e2}).\\
It must be emphasized that equations (\ref{e1}) and (\ref{e2}) in
particular bring out apart from the random feature a holistic or
Machian feature in which the large scale universe and the micro
world are inextricably tied up, as against the usual reductionist
view discussed in detail earlier. This is in fact inescapable if we
are to consider a Brownian Heap. This interpretation in which the
extent $R (\mbox{or}\, T)$ in (\ref{e1}) (or (\ref{e2})) is a
dispersion also explains the fractal dimensionality 2: If the steps
were laid out one beside the other unidirectionally as in
conventional thinking, then we would have the usual dimensionality
one. For, instead of (\ref{e1}), we would have,
$$R = N l$$
This again is tied up with a model in terms of a Weiner process (a Random
Walk), as we will see below.\\
There is another nuance. Newtonian space was a passive container
which "contained" matter and interactions - these latter were actors
performing on the fixed platform of space. But our view is in the
spirit of Liebniz \cite{lucas} for whom the container of space was
made up of the contents - the actors, as it were, made up the stage
or platform. This also implies the background independence alluded
to earlier, a feature shared by General Relativity.\\
It should also be observed that the cut off length for fractal
behaviour depends on the mass, via the de Broglie or Compton
wavelength. The de Broglie wavelength is the non-relativistic
version of the Compton wavelength. Indeed as has been shown in
detail \cite{ijpap,cu}, it is the zitterbewegung or self-interaction
effects within the minimum cut off Compton wavelength that give rise
to the inertial mass. So the appearance of mass in the minimum cut
off Compton (or de Broglie) scale is quite natural. This point will
be
analyzed further in the sequel.\\
We can appreciate that the fractal nature and a stochastic
underpinning are interrelated: for scales less than the Compton (or
de Broglie) wavelength, time is irregular and can be modelled by a
double Wiener process\cite{r68}. This will be shown to lead to the
complex wave function of Quantum Mechanics, which is one of its
distinguishing characteristics (in contrast to
Classical theory where complex quantities are a mathematical artifice).\\
To appreciate all this let us consider the motion of a particle with
position given by $x(t)$, subject to random correction given by, as
in the usual theory, (Cf.\cite{r27,r47,r67}),
$$|\Delta x| = \sqrt{<\Delta x^2 >} \approx \nu \sqrt{\Delta t},$$
\begin{equation}
\nu = \hbar/m, \nu \approx l v\label{e3}
\end{equation}
where $\nu$ is the so called diffusion constant and is related to
the mean free path $l$ as above. We can then proceed to deduce the
Fokker-Planck equation as follows (Cf.[ref.\cite{r27} for
details):\\
We first define the forward and backward velocities corresponding to
having time going forward and backward (or positive or negative time
increments) in the usual manner,
\begin{equation}
\frac{d_+}{dt} x (t) = {\bf b_+} \, , \, \frac{d_-}{dt} x(t) = {\bf
b_-}\label{ex1}
\end{equation}
This leads to the Fokker-Planck equations
$$
\partial \rho / \partial t + div (\rho {\bf b_+}) = V \Delta \rho
,$$
\begin{equation}
\partial \rho / \partial t + div (\rho {\bf b_-}) = - U \Delta
\rho\label{ex2}
\end{equation}
defining
\begin{equation}
V = \frac{{\bf b_+ + b_-}}{2} \quad ; U = \frac{{\bf b_+ - b_-}}{2}
\label{ex3}
\end{equation}
We get on addition and subtraction of the equations in (\ref{ex2})
the equations
\begin{equation}
\partial \rho / \partial t + div (\rho V) = 0\label{ex4}
\end{equation}
\begin{equation}
U = \nu \nabla ln\rho\label{ex5}
\end{equation}
It must be mentioned that $V$ and $U$ are the statistical averages
of the respective velocities. We can then introduce the definitions
\begin{equation}
V = 2 \nu \nabla S\label{ex6}
\end{equation}
\begin{equation}
V - \imath U = -2 \imath \nu \nabla (l n \psi)\label{ex7}
\end{equation}
The decomposition of the Schrodinger wave function as
$$\psi = \sqrt{\rho} e^{\imath S/\hbar}$$
leads to the well known Hamilton-Jacobi type equation
\begin{equation}
\frac{\partial S}{\partial t} = -\frac{1}{2m} (\partial S)^2 + \nu
+Q,\label{e4}
\end{equation}
where
$$Q = \frac{\hbar^2}{2m} \frac{\nabla^2 \sqrt{\rho}}{\sqrt{\rho}}$$
From (\ref{ex6}) and (\ref{ex7}) we can finally deduce the usual
Schrodinger equation or (\ref{e4}) \cite{r68}.\\
We note that in this formulation three conditions are assumed,
conditions whose import has not been clear. These are \cite{r27}:\\
(1) The current velocity is irrotational. Thus, there exists a
function $S(x,t)$ such that
$$m \vec V = \vec \nabla S$$
(2) In spite of the fact that the particle is subject to random
alterations in its motion there exists a conserved energy, defined
in terms of its
probability distribution.\\
(3) The diffusion constant is inversely proportional to the inertial
mass of the particle, with the constant of proportionality being a
universal constant $\hbar$ (Cf. equation (\ref{e3})):
$$\nu = \frac{\hbar}{m}$$
We note that the complex feature above disappears if the fractal or
non-differential character is not present, (that is, the forward and
backward time derivatives(\ref{ex3}) are equal): Indeed the fractal
dimension 2 also leads to the real coordinate becoming complex. What
distinguishes Quantum Mechanics is the adhoc feature, the diffusion
constant $\nu$ of (\ref{e3}) in Nelson's theory and the "Quantum
potential" $Q$ of
(\ref{e4}) which appears in Bohm's theory as well, though with a different meaning.\\
Interestingly from the Uncertainty Principle,
$$m \Delta x  \frac{\Delta x}{\Delta t} \sim \hbar$$
we get back equation (\ref{e3}) of Brownian motion. This shows the
close connection on the one hand, and provides, on the other hand, a
rationale for the particular, otherwise adhoc identification of
$\nu$ in (\ref{e3}) - its being proportional to $\hbar$.\\
We would like to emphasize that we have arrived at the Quantum
Mechanical Schrodinger equation from Classical considerations of
diffusion, though with some new assumptions. In the above,
effectively we have introduced a complex velocity $V - \imath U$
which alternatively means that the real coordinate $x$ goes into a
complex coordinate
\begin{equation} x \to x + \imath x'\label{De9d}
\end{equation}
To see this in detail, let us rewrite (\ref{ex3}) as
\begin{equation}
\frac{dX_r}{dt} = V, \quad \frac{dX_\imath}{dt} = U,\label{De10d}
\end{equation}
where we have introduced a complex coordinate $X$ with real and
imaginary parts $X_r$ and $X_\imath$, while at the same time using
derivatives with respect
to time as in conventional theory.\\
We can now see from (\ref{ex3}) and (\ref{De10d}) that
\begin{equation}
W = \frac{d}{dt} (X_r - \imath X_\imath )\label{De11d}
\end{equation}
That is, in this non relativistic development either we use forward
and backward time derivatives and the usual space coordinate as in
(\ref{ex3}), or we use the derivative with respect to the usual time
coordinate but introduce complex space coordinates as in
(\ref{De9d}).\\
Let us briefly analyze this aspect though we will return to it
later. To bring out the new input here, we will consider the
diffusion equation (\ref{e3}) in only one dimension for the moment.
We note that through (\ref{ex3}) we have introduced a complex
velocity $W$, as indeed can be seen from (\ref{De10d}) and
(\ref{De11d}) as well. Furthermore (\ref{ex5}) and (\ref{ex6}) show
that both $U$ and $V$ can be written as gradients in the form
$$\vec{V} = \vec{\nabla} f$$
\begin{equation}
\vec{U} = \vec{\nabla} g\label{EA}
\end{equation}
Furthermore the equation of continuity, (\ref{ex4}) shows that for
nearly constant and homogenous density $\rho$ we have
\begin{equation}
\vec{\nabla} \cdot \vec{V} = 0\label{EB}
\end{equation}
where we are still retaining the vector notation. This implies that
$f$ and so also $g$ satisfy the Laplacian equation
\begin{equation}
\nabla^2 f = 0\label{EC}
\end{equation}
In this case given (\ref{EC}),it is well known from the Theory of
Fluid flow \cite{joos} that the trajectories $f = \mbox{constant}$
and $g = \mbox{constant}$ are orthogonal, with, in the case of
spherical symmetry, the former representing radial stream lines and
the latter circles around the origin (or more generally closed
curves). We also see that (\ref{EB}) shows that the velocity is
solenoidal, and $\vec{V}$ being a gradient, by (\ref{ex6}), also
irrotational. We would then expect that the circulation given by the
expression
\begin{equation}
\Gamma = m \oint \vec{V} \cdot d\vec{s}\label{ED}
\end{equation}
would vanish. All this is true in a simply connected space. However
if the space is multiply connected, the origin being the
singularity, then the circulation (\ref{ED}) does not vanish. We
argue that this is the Quantum Mechanical spin, and will return to
this point. But briefly, $\Gamma$ in (\ref{ED}) equals the Quantum
Mechanical spin $h/2$. This follows, if we take the radius of the
circuit of integration to be the Compton wavelength $\hbar/mc$ and
remember that at this distance, the velocity equals $c$.\\
The interesting thing is that starting from a single real
coordinate, we have ended up with a complex coordinate, and have
characterized thereby, the Quantum Mechanical spin. Indeed as we
will shortly see it was noticed by Newman in the derivation of the
Kerr-Newman metric, that an inexplicable imaginary shift gives
Quantum Mechanical spin. In other words Quantum Mechanics results
from a complexification of coordinates, this as can be seen now,
being symptomatic of multiply connected spaces, and modelled by the
Weiner process
above.\\
Finally, it may be remarked that the original Nelsonian theory
itself has been criticized by different scholars
\cite{A1}-\cite{A5}.\\
To get further insight into the foregoing considerations, let us
start with the Langevin equation in the absence of external
forces,\cite{r47,r69}
$$m \frac{dv}{dt} = -\alpha v + F'(t)$$
where the coefficient of the frictional force is given by Stokes's
Law \cite{joos}
$$\alpha = 6\pi \eta a$$
$\eta$ being the coefficient of viscosity, and where we are
considering a
sphere of radius $a$. This then leads to two cases.\\
Case (i):\\
For $t$, there is a cut off time $\tau$. It is known (Cf.\cite{r47})
that there is a characteristic time constant of the system, given by
$$\frac{m}{\alpha} \sim \frac{m}{\eta a},$$
so that, from Stokes's Law, as
$$\eta = \frac{mc}{a^2} \, \mbox{or}\, m = \eta \frac{a^2}{c}$$
we get
$$\tau \sim \frac{ma^2}{mca} = \frac{a}{c},$$
that is $\tau$ is the Compton time.\\
The expression for $\eta$ which follows from the fact that
$$F_x = \eta (\Delta s) \frac{dv}{dz} = m\dot v = \eta \frac{a^2}{c} \dot v ,$$
shows that the intertial mass is due to a type of "viscosity" of the
background Zero Point Field (ZPF). (Cf. also ref.\cite{r70}).\\
To sum up case (i), if there is a cut off $\tau$, the stochastic
formulation leads us back to the
minimum space time intervals $\sim$ Compton scale.\\
To push these small scale considerations further, we have, using the
Beckenstein radiation equation\cite{r72},
$$t \equiv \tau = \frac{G^2m^3}{\hbar c^4} = \frac{m}{\eta a} = \frac{a}{c}$$
which gives
$$a = \frac{\hbar}{mc} \quad \mbox{if} \quad \frac{Gm}{c^2} = a$$
In other words the Compton wavelength equals the Schwarzchild
radius, which automatically gives us the Planck mass. Thus as noted
the inertial mass is thrown up in these considerations. We will also
see that the Planck mass leads to other particle
masses.\\
On the other hand if we work with $t \geq \tau$ we get
$$ac = \frac{2kT}{\eta a}$$
whence
$$kT \sim mc^2,$$
which is the Hagedorn formula for Hadrons\cite{r73}.\\
Thus both the Planck scale and the Compton wavelength Hadron scale
considerations follow meaningfully.\\
Case (ii):\\
If there is no cut off time $\tau$, as is known, we get back,
equation (\ref{e3}),
$$\Delta x = \nu \sqrt{\Delta t}$$
and thence Nelson's derivation of the non relativistic Schrodinger
equation. We can see here that the absence of a space time cut off
leads to the non-relativistic theory, but on the contrary the cut off leads to the
origin of the inertial mass (and as we will see, relativity itself). On the other
hand, as we saw, the cut off is symptomatic of a multiply connected space- where
we cannot shrink circuits to a point.\\
The relativistic generalization of the above considerations to the
Klein-Gordon equation has been even more troublesome\cite{r8}. In
this case, there are further puzzling features apart from the
luminal velocity as in the Dirac equation. For Lorentz invariance, a
discrete time is further required. Interestingly, as we will see
Snyder had shown that discrete spacetime is compatible with Lorentz
transformations. Here again, the Compton wavelength and time cut off
 will be seen to make the whole picture transparent.\\
The stochastic derivation of the Dirac equation introduces a further
complication. There is a spin reversal with the frequency
$mc^2/\hbar$. This again is readily explainable in the earlier
context of zitterbewegung in terms of the Compton time.
Interestingly the resemblance of such a Weiner process to the
zitterbewegung
of the electron was noticed by Ichinose\cite{r74}.\\
Thus in all these cases once we recognize that the Compton
wavelength and time
are minimum cut off intervals, the obscure or adhoc features become meaningful.\\
We would like to reiterate that the origin of the Compton wavelength
is the random walk equation (\ref{e1})! One could then argue that
the Compton time (or Chronon) automatically follows. This was shown
by Hakim \cite{r14,r16}. Intutively, we can see that a discrete
space would automatically imply discrete time. For, if $\Delta t$
could $\to 0$, then all velocities, $lim_{\Delta t \to 0}
|\frac{\Delta x}{\Delta t}|$ would $\to \infty \,\mbox{as}\, |\Delta
x |$ does not tend to $0!$ So there would be a minimum time cut off
and a maximal velocity and this in conjunction with symmetry
considerations can be taken to be the basis of special relativity as
we will see below in more
detail.\\
In fact one could show that quantized spacetime is more fundamental
than quantized energy and indeed would lead to the latter. To put it
simply the frequency is given by $c/\lambda$, where $\lambda$ the
wavelength is itself discrete and hence so also is the frequency.
One could then deduce Planck's law as will be seen in the next
Section (Cf.\cite{r75}). This of course, is the starting point of
Quantum
Theory itself.\\
At this stage we remark that in the case of the Dirac electron, the
point electron has the velocity of light and is subject to
zitterbewegung within the Compton wavelength. The thermal wavelength
for such a motion is given by
$$\lambda = \sqrt \frac{\hbar^2}{mkT} \sim \mbox{De Broglie \quad wave length}$$
by virtue of the fact that now $kT \sim mv^2$ itself. In the limit
$v \to c$ in the spirit of the luminal velocity of the point Dirac
electron or, using the earlier relation, $kT \sim mc^2$, $\lambda$
becomes the Compton wavelength. To look at this from another point
of view, it is known that for a collection of relativistic
particles, the various mass centres form a two-dimensional disc
perpendicular to the angular momentum vector $\vec L$ and with
radius (ref.\cite{r32})
\begin{equation}
r = \frac{L}{mc}\label{e29}
\end{equation}
Further if the system has positive energies, then it must have an
extension greater than $r$, while at distances of the order of
$r$ we begin to encounter negative energies.\\
If we consider the system to be a particle of spin or angular
momentum $\frac{\hbar}{2}$, then equation (\ref{e29}) gives, $r =
\frac{\hbar}{2mc}$. That is we get back the Compton wavelength.
Another interesting feature which we will encounter later is the two
dimensionality of the space or disc
of mass centres.\\
On the other hand it is known that, if a Dirac particle is
represented by a Gausssian packet, then we begin to encounter
negative energies precisely at the same  Compton wavelength as
above. These considerations show the interface between
Classical and Quantum considerations.\\
Infact as has been shown it is this circumstance that leads to
inertial mass, while gravitation and electromagnetism (as for
example brought out by the Kerr-Newman metric) and indeed QCD
interactions also will be seen to follow. In the light of the above
remarks, it appears that the fractal or Brownian Heap character of
space time is at the root of Quantum behaviour.
\section{Spacetime}
As remarked in the previous section, the fact that forward and
backward time derivatives in the double Wiener process do not cancel
leads to a complex velocity (cf.\cite{r68}), $V-\imath U$. That is,
the usual space coordinate $x$ (in one dimension for simplicity) is
replaced by a coordinate like $x+\imath x'$, where $x'$ is a non
constant function of time that is, a new imaginary coordinate is
introduced. We will now
show that it is possible to consistently take $x' = ct$.\\
Let us take the simplest choice for $x'$, viz., $x' = \lambda t$.
Then the imaginary part of the complex velocity in (\ref{De11d})is
given by $U = \lambda$. Then we have (cf.\cite{r67}),
$$U = \nu \frac{d}{dx}(ln\rho) = \lambda$$
where $\nu$ and $\rho$ have been defined in (\ref{e3}), and in the
equation leading to (\ref{e4}). We thus have, $\rho = e^{\gamma x}$,
where $\gamma = \lambda/\nu$ and the quantum potential of (\ref{e4})
is given by
\begin{equation}
Q \sim \frac{\hbar^2}{2m}\cdot \quad \gamma^2\label{e5}
\end{equation}
In this stochastic formulation with Compton wavelength cut off, it
is known that $Q$ turns out to be the inertial energy
$mc^2$. It then follows from (\ref{e5}) and the definition of
$\gamma$, that $\lambda \approx c$.\\
In other words it is in the above stochastic formulation that we see
the emergence of the spacetime coordinates $(x,\imath ct$) and
Special Relativity from a Weiner process in which time is a back and
forth process. All this has been in one dimension.\\
If we now generalize to three spatial dimensions, then as we will
see in a moment \cite{r76}, we get the quarternion formulation with
the three Pauli spin matrices replacing $\imath$, giving the purely
Quantum Mechanical spin half of Dirac. On the other hand, the above
formulation with minimum space time cut offs will also be shown to
lead independently to the Dirac equation. Thus the origin of special
relativity, inertial mass and the Quantum Mechanical spin half is
the minimum
space time cut offs.\\
We digress for a moment to observe that equations (\ref{e1}) and
(\ref{e2}) indicate that the Compton scale is a fundamental unit of
space time. We will now show that this quantized space time leads to
Planck's quantized energy, as was
briefly seen in the previous section.\\
The derivation is similar
to the well known theory\cite{r77}.\\
Let the energy be given by
$$E = g(\nu)$$
Then, $f$ the average energy associated with each mode is given by,
$$f = \frac{\sum_\nu g(\nu) e^{-g(\nu)/kT}}{\sum_\nu e^{-g(v)/kT}}$$
Again, as in the usual theory, a comparison with Wien's functional
relation, gives,
$$f = \nu F (\nu/kT),$$
whence,
$$E = g(\nu)\propto \nu,$$
which is Planck's law.\\
Yet another way of looking at it is, as the momentum and frequency
of the
classical oscillator have discrete spectra so does the energy.\\
\section{Further Considerations}
To see all this in greater detail, we observe that if we treat an
\index{electron}electron as a \index{Kerr-Newman}Kerr-Newman
\index{black hole}black hole, then we get the correct Quantum
Mechanical $g=2$ factor, but the horizon of the  \index{black
hole}black hole becomes complex \cite{cu,r40}.
\begin{equation}
r_+ = \frac{GM}{c^2} + \imath b, b \equiv (\frac{GQ^2}{c^4} + a^2 -
\frac{G^2M^2}{c^4})^{1/2}\label{De1}
\end{equation}
$G$ being the \index{gravitation}gravitational constant, $M$ the
\index{mass}mass and $a \equiv L/Mc,L$ being the angular momentum.
While (\ref{De1}) exhibits a naked singularity, and as such has no
physical meaning, we note that from the realm of Quantum Mechanics
the position coordinate for a \index{Dirac}Dirac particle in
conventional theory is given by
\begin{equation}
x = (c^2p_1H^{-1}t) + \frac{\imath}{2} c\hbar (\alpha_1 -
cp_1H^{-1})H^{-1}\label{De2}
\end{equation}
an expression that is very similar to (\ref{De1}). Infact as was argued in detail
\cite{cu} the imaginary parts of both (\ref{De1}) and (\ref{De2}) are the same, being
of the order of the \index{Compton wavelength}Compton wavelength.\\
It is at this stage that a proper physical interpretation begins to emerge.
\index{Dirac}Dirac himself observed as noted, that to interpret (\ref{De2})
meaningfully, it must be remembered that Quantum Mechanical measurements are really
averaged over the \index{Compton scale}Compton scale: Within the scale there are the
unphysical \index{zitterbewegung}zitterbewegung effects: for a point
\index{electron}electron the velocity equals that of light.\\
Once such a minimum \index{spacetime}spacetime scale is invoked,
then we have a non commutative geometry as shown by
\index{Snyder}Snyder more than fifty years ago \cite{sny1,sny2}:
$$[x,y] = (\imath a^2/\hbar )L_z, [t,x] = (\imath a^2/\hbar c)M_x, etc.$$
\begin{equation}
[x,p_x] = \imath \hbar [1 + (a/\hbar )^2 p^2_x];\label{De3}
\end{equation}
The relations (\ref{De3}) are compatible with \index{Special
Relativity}Special Relativity. Indeed such minimum
\index{spacetime}spacetime models were studied for several decades,
precisely to overcome the \index{divergences}divergences encountered
in \index{Quantum Field Theory}Quantum Field Theory
\cite{cu},\cite{sny2}-\cite{fink},
\cite{x3,x9}.\\
Before proceeding further, it may be remarked that when the square of a, which we
will take to be the \index{Compton wavelength}Compton wavelength (including the
\index{Planck scale}Planck scale, which is a special case of the
\index{Compton scale}Compton scale for a \index{Planck}Planck \index{mass}mass viz.,
$10^{-5}gm$), in view of the above comments  can be neglected, then we return to point
\index{Quantum Theory}Quantum Theory.\\
It is interesting that starting from the \index{Dirac}Dirac
coordinate in (\ref{De2}), we can deduce the non commutative
geometry (\ref{De3}), independently. For this we note that the
$\alpha$'s in (\ref{De2}) are given by
$$\vec \alpha = \left[\begin{array}{ll}
\vec \sigma \quad 0\\
0 \quad \vec \sigma
\end{array}
\right]\quad \quad ,$$
the $\sigma$'s being the \index{Pauli}Pauli
matrices. We next observe that the first term on the right hand side
is the usual Hermitian position. For the second term which contains
$\alpha$, we can easily verify from the commutation relations of the
$\sigma$'s that
\begin{equation}
[x_\imath , x_j] = \beta_{\imath j} \cdot l^2\label{DeA}
\end{equation}
where $l$ is the \index{Compton scale}Compton scale.\\
There is another way of looking at this. Let us consider the one
dimensional coordinate in (\ref{De2}) or (\ref{De1}) to be complex.
We now try to generalize this \index{complex coordinate}complex
coordinate to three dimensions. Then as briefly noted, in the
previous Section, we encounter a surprise - we end up with not
three, but four dimensions,
$$(1, \imath) \to (I, \sigma),$$
where $I$ is the unit $2 \times 2$ matrix. We get the special
relativistic \index{Lorentz}Lorentz invariant metric at the same
time. (In this sense, as noted by Sachs \cite{r76}, Hamilton who
made this generalization would have hit upon \index{Special
Relativity}Special Relativity, if he had identified the new fourth
coordinate
with time).\\
That is,\\
$$x + \imath y \to Ix_1 + \imath x_2 + jx_3 + kx_4,$$
where $(\imath ,j,k)$ now represent the \index{Pauli}Pauli matrices;
and, further,
$$x^2_1 + x^2_2 + x^2_3 - x^2_4$$
is invariant. Before proceeding further, we remark that special
relativistic time emerges above from the generalization of the
complex one dimensional space coordinate to three dimensions, just
as the relativistic time came out of the one dimensional
space coordinate as seen earlier.\\
While the usual \index{Minkowski}Minkowski four vector transforms as the basis of the
four dimensional representation of the \index{Poincare}Poincare group, the two dimensional
representation of the same group, given by the right hand side in terms of
\index{Pauli}Pauli matrices, obeys the quaternionic algebra of the second rank
\index{spin}spinors (Cf.Ref.\cite{sakharov,bgsfpl,r76} for details).\\
To put it briefly, the \index{quarternion}quarternion number field
obeys the group property and this leads to a number system of
quadruplets as a minimum extension. In fact one representation of
the two dimensional form of the \index{quarternion}quarternion basis
elements is the set of \index{Pauli}Pauli matrices. Thus a
\index{quarternion}quarternion may be expressed in the form
$$Q = -\imath \sigma_\mu x^\mu = \sigma_0x^4 - \imath \sigma_1 x^1 - \imath \sigma_2 x^2 - \imath \sigma_3 x^3 = (\sigma_0 x^4 + \imath \vec \sigma \cdot \vec r)$$
This can also be written as
$$Q = -\imath \left(\begin{array}{ll}
\imath x^4 + x^3 \quad x^1-\imath x^2\\
x^1 + \imath x^2 \quad \imath x^4 - x^3
\end{array}\right).$$
As can be seen from the above, there is a one to one correspondence between a
\index{Minkowski}Minkowski four-vector and $Q$. The invariant is now given by
$Q\bar Q$, where $\bar Q$ is the complex conjugate of $Q$.\\
However, as is well known, there is a lack of
\index{spacetime}spacetime reflection \index{symmetry}symmetry in
this latter formulation. If we require reflection
\index{symmetry}symmetry also, we have to consider the four
dimensional representation,
$$(I, \vec \sigma) \to \left[\left(\begin{array}{ll}
I \quad 0 \\
0 \quad -I
\end{array}\right), \left(\begin{array}{ll}
0 \quad \vec \sigma \\
\vec \sigma \quad 0
\end{array}\right)\right] \equiv  (\Gamma^\mu)$$
(Cf.also.ref. \cite{rr87} for a detailed discussion). The motivation for such a reflection
\index{symmetry}symmetry is that usual laws of physics, like \index{electromagnetism}
electromagnetism do indeed show the symmetry.\\
We at once deduce \index{spin}spin and \index{Special
Relativity}Special Relativity and
 the geometry (\ref{De3}) in these considerations. This is a transition that has
 been long overlooked \cite{br2}.
 It must also be mentioned that \index{spin}spin half itself is relational and refers to
 three dimensions, to a \index{spin}spin network infact \cite{ar9,r40}. That is,
 \index{spin}spin half is not meaningful in a single particle \index{Universe}Universe.\\
While a relation like (\ref{DeA}) above has been in use recently, in
non commutative models, we would like to stress that it has been
overlooked that the origin of this non commutativity
lies in the original \index{Dirac}Dirac coordinates.\\
The above relation shows on comparison with the position-momentum
commutator that the coordinate $\vec x$ also behaves like a
``momentum''. This can be seen directly from the \index{Dirac}Dirac
theory itself where we have \cite{r12}
\begin{equation}
c\vec \alpha = \frac{c^2\vec p}{H} - \frac{2\imath}{\hbar} \hat x
H\label{Dea}
\end{equation}
In (\ref{Dea}), the first term is the usual momentum. The second term is the extra
``momentum'' $\vec p$ due to \index{zitterbewegung}zitterbewegung.\\
Infact we can easily verify from (\ref{Dea}) that
\begin{equation}
\vec p = \frac{H^2}{\hbar c^2}\hat x\label{Deb}
\end{equation}
where $\hat x$ has been defined in (\ref{Dea}).\\
We finally investigate what the angular momentum $\sim \vec x \times
\vec p$ gives - that is, the angular momentum at the \index{Compton
scale}Compton scale. We can easily show that
\begin{equation}
(\vec x \times \vec p)_z = \frac{c}{E} (\vec \alpha \times \vec p)_z
= \frac{c}{E} (p_2\alpha_1 - p_1\alpha_2)\label{Dec}
\end{equation}
where $E$ is the eigen value of the Hamiltonian operator $H$.
Equation (\ref{Dec}) shows that the usual angular momentum but in
the context of the minimum \index{Compton scale}Compton scale cut
off, leads to the ``mysterious'' Quantum Mechanical \index{spin}spin.\\
In the above considerations, we started with the \index{Dirac}Dirac
equation and deduced the underlying non commutative geometry of
spacetime. Interestingly, starting with Snyder's non commutative
geometry, based solely on \index{Lorentz}Lorentz invariance and a
minimum \index{spacetime}spacetime length, which we have taken to be
the \index{Compton scale}Compton scale, (\ref{De3}),
it is possible to deduce the relations (\ref{Dec}), (\ref{Deb}) and the
\index{Dirac}Dirac equation itself as we will see later.\\
We have thus established the correspondence between considerations starting from
the \index{Dirac}Dirac theory of the \index{electron}electron and \index{Snyder}Snyder's
(and subsequent) approaches based on a minimum \index{spacetime}spacetime interval
and \index{Lorentz}Lorentz covariance. It can be argued from an alternative point
of view that \index{Special Relativity}Special Relativity operates outside the
\index{Compton wavelength}Compton wavelength as we saw earlier.\\
We started with the \index{Kerr-Newman}Kerr-Newman  \index{black
hole}black hole. Infact the derivation of the
\index{Kerr-Newman}Kerr-Newman  \index{black hole}black hole itself
begins with a Quantum Mechanical \index{spin}spin yielding complex
shift, which Newman has found inexplicable even after several
decades \cite{newman1,newman2}. As he observed, "...one does not
understand why it works. After many years of study I have come to
the conclusion that it works simply by accident". And again, "Notice
that the magnetic moment $\mu = ea$ can be thought of as the
imaginary part of the charge times the displacement of the charge
into the complex region... We can think of the source as having a
complex center of charge and that the magnetic moment is the moment
of charge about the center of charge... In other words the total
complex angular momentum vanishes around any point $z^a$ on the
complex world-line. From this complex point of view the spin angular
momentum is identical to orbital, arising from an imaginary shift of
origin rather than a real one... If one again considers the particle
to be "localized" in the sense that the complex center of charge
coincides with the complex center of mass, one again obtains the
Dirac gyromagnetic ratio..."\\
The unanswered question has been, why does a complex shift somehow
represent \index{spin}spin about that axis? The question has now
been answered. Complexified spacetime is symptomatic of fuzzy
spacetime and a non commutative geometry and Quantum Mechanical
\index{spin}spin and relativity. Indeed Zakrzewski has shown in a
classical context that non commutativity implies \index{spin}spin
\cite{zak,bgschubykalo}. We will return to these considerations
later.\\
The above considerations recovered the Quantum Mechanical
\index{spin}spin together with classical relativity, though the
price to pay for this was minimum spacetime intervals and
\index{noncommutative}noncommutative geometry.
\section{The Path Integral Formulation}
We come to another description of Quantum Mechanics and first argue
that the alternative Feynman Path Integral formulation essentially
throws up fuzzy spacetime. To recapitulate \cite{fh,nottale,it}, if
a path is given by
$$x = x(t)$$
then the probability amplitude is given by
$$\phi (x) = e^{\imath \int^{t_2}_{t_1}L(x, \dot{x})dt}$$
So the total probability amplitude is given by
$$\sum_{x(t)} \phi (x) = \sum e^{\imath \int^{t_2}_{t_1}L(x,\dot{x})dt} \equiv \sum e^{\frac{\imath}{\hbar}S}$$
In the Feynman analysis, the path
$$x = \bar{x} (t)$$
appears as the actual path for which the action is stationery. From a physical
point of view, for paths very close to this, there is constructive interference, whereas
for paths away from this the interference is destructive.\\
We will see later that this is in the spirit of the formulation of
the random phase. However it is well known that the convergence of
the integrals requires the Lipshitz condition viz.,
\begin{equation}
\Delta x^2 \approx a \Delta t\label{e1b}
\end{equation}
We could say that only those paths satisfying (\ref{e1b})
constructively interfere. We would now like to observe that
(\ref{e1b}) is the same as the Brownian or Diffusion equation (3)
related to our earlier discussion of the Weiner process. The point
is that (\ref{e1b}) again implies a minimum spacetime cut off, as
indeed was noted by Feynman himself \cite{fh}, for if
$\Delta t$ could $\to 0$, then the velocity would $\to \infty$.\\
To put it another way we are taking averages over an interval
$\Delta t$, within which there are unphysical processes as noted. It
is only after the average is taken, that we recover physical
spacetime intervals which hide the fractality or unphysical feature.
If in the above, $\Delta t$ is taken as the Compton time (and $a$ is
identified with the earlier $\nu$, then we recover for the root mean
squared
velocity, the velocity of light.\\
As we have argued in detail this is exactly the situation which we encounter in the
Dirac theory of the electron. There we have the unphysical zitterbewegung effects
within the Compton time $\Delta t$ and as $\Delta t \to 0$ the velocity of the
electron tends to the maximum possible velocity, that of light \cite{dirac}.
 It is only after averaging over the Compton scale that we recover meaningful physics.\\
This  existence of a minimum spacetime scale, it has been argued is
the origin of fuzzy spacetime, described by a noncommutative
geometry, consistent with Lorentz invariance viz., equations
(\ref{De3}) and (\ref{DeA}).\\
We reiterate that the momentum position commutation relations lead
to the usual Quantum Mechanical commutation relations in the usual
(commutative) spacetime if $O(l^2)$ is neglected where $l$ defines
the minimum scale. Indeed, we have at the smallest scale, a quantum
of area reflecting the fractal dimension, the Quantum Mechanical
path having the fractal dimension 2 (Cf.ref.\cite{r50}). It is this
``fine structure'' of spacetime which is expressed in
 the noncommutative structure (\ref{De3}) or (\ref{DeA}). Neglecting $O(l^2)$ is equivalent to
 neglecting the above and returning to usual spacetime. In other words Snyder's
 purely classical considerations at a Compton scale lead to Quantum Mechanics.\\
In the light of the above comments, we can now notice that within
the Compton time, we have a double Weiner process leading to non
differentiability with respect to time. That is, at this level time
in our usual sense does not exist. To put it another way, within the
Compton scale we have the complex or non-Hermitian position
coordinates for the Dirac electron and zitterbewegung effects -
these are unphysical, non local and chaotic in a literal sense.\\
This is a Quantum Mechanical and an experimental fact. It expresses
the Heisenberg Uncertainty Principle - space time points imply
infinite momenta and energies and are thus not meaningful
physically. However as noted earlier Quantum Theory has lived with
this contradiction. To put it simply to measure space or time
intervals we need units which can be to a certain extent and not
indefinitely subdivided - but already this is the origin of
discreteness. That is, our measurements are resolution dependent. So
physical time emerges at values greater than the minimum unit, which
has been shown to be at the Compton scale. Going to the limit of
space-time points leads to the well known infinities of Quantum
Field Theory (and classical
electron theory) which require renormalization for their removal.\\
The conceptual point here is that time is in a sense synonymous with
change, but this change has to be tractable or physical. The non
differentiability with respect to time, symbolized and modeled by
the double Weiner process, within the Compton time, precisely
highlights time or change which is not tractable, that is is
unphysical. However Physics, tractability and differentiability
emerge from this indeterminism once averages over the zitterbewegung
or Compton scale are taken. It is now possible to track time
physically in
terms of multiples of the Compton scale.\\
\section{Discussion}
1. We would like to make the following observations:\\
i) We have in effect equated the statistical fluctuations, when
there are $N$ particles to the Quantum Mechanical fluctuations. The former
fluctuations take place over a scale $\sim R/\sqrt{N}$, where $R$ is the
size of the system of particles and $N$ is the number of particles in the system.
The Quantum Mechanical fluctuations take place at a scale of the order of
the Compton wavelength. Apart from the fact that the equality of these two
has been taken to be an empirical coincidence, we actually deduce this equality
in our cosmology in the next Chapter. Thus the equality is no longer
accidental or ad hoc. However a nuance must be borne in mind. In the conventional
theory, the Quantum Mechanical fluctuation is a reductionist effect, whereas
the statistical fluctuation is a "thermodynamic" or statistical effect in a
collection of particles.\\
ii) In the random mechanical approach, including Nelson's, we
encounter "potential" $Q$- this represents in the usual theory a
peculiar correlation between the random motion of a particle and its
probability distribution function.\\
iii) We would like to point out that it would be reasonable to
expect that the Weiner process discussed earlier is related to the
ZPF which is the Zero Point Energy of a Quantum Harmonic oscillator.
We can justify this expectation as follows: Let us denote the
forward and backward time derivatives as before by $d_+$ and $d_-$.
In usual theory where time is differentiable, these two are equal,
but we have on the contrary taken them to be unequal. Let
\begin{equation}
d_- = a - d_+\label{ae}
\end{equation}
Then we have from Newton's second law in the absence of forces,
\begin{equation}
\ddot{x} + k^2 x = a\dot{x}\label{be}
\end{equation}
wherein the new nondifferentiable effect (\ref{ae}) is brought up.
In a normal vacuum with usual derivatives and no external forces,
Newtonian Mechanics would give us instead the equation
\begin{equation}
\ddot{x} = 0\label{ce}
\end{equation}
A comparison of (\ref{be}) and (\ref{ce}) shows that the Weiner
process converts a uniformly moving particle, or a particle at rest
into an oscillator. Indeed in (\ref{be}) if we take as a first
approximation
\begin{equation}
\dot{x} \approx \langle \dot{x} \rangle = 0\label{de}
\end{equation}
then we would get the exact oscillator equation
\begin{equation}
\ddot{x} + k^2 x = 0\label{ee}
\end{equation}
for which in any case, consistently (\ref{de}) is correct. We can
push these considerations even further and deduce alternatively, the
Schrodinger equation, as seen earlier. The genesis of Special
Relativity too can be found in the Weiner process. Let us examine this
more closely.\\
We first define a complete set of base states by the subscript
$\imath \quad \mbox{and}\quad U (t_2,t_1)$ the time elapse operator
that denotes the passage of time between instants $t_1$ and $t_2$,
$t_2$ greater than $t_1$. We denote by, $C_\imath (t) \equiv <
\imath |\psi (t)
>$, the amplitude for the state $|\psi (t) >$ to be in the state $|
\imath >$ at time $t,$ and \cite{ijpap,cu}
$$< \imath |U|j > \equiv U_{\imath j}, U_{\imath j}(t + \Delta t,t) \equiv
\delta_{\imath j} - \frac{\imath}{\hbar} H_{\imath j}(t)\Delta t.$$
We can now deduce from the super position of states principle that,
\begin{equation}
C_\imath (t + \Delta t) = \sum_{j} [\delta_{\imath j} -
\frac{\imath} {\hbar} H_{\imath j}(t)\Delta t]C_j (t)\label{xe}
\end{equation}
and finally, in the limit,
\begin{equation}
\imath \hbar \frac{dC_\imath (t)}{dt} = \sum_{j} H_{\imath
j}(t)C_j(t)\label{fe}
\end{equation}
where the matrix $H_{\imath j}(t)$ is identified with the
Hamiltonian operator. We have argued earlier at length that
(\ref{fe}) leads to the Schrodinger equation \cite{ijpap,cu}. In the
above we have taken the usual unidirectional time to deduce a non
relativistic Schrodinger equation. If however we consider a Weiner
process in (\ref{xe}) then we will have to consider instead of
(\ref{fe})
\begin{equation}
C_\imath (t - \Delta t) - C_\imath (t + \Delta t) = \sum_{j}
\left[\delta_{\imath j} - \frac{\imath}{\hbar} H_{\imath
j}(t)\right] C_j^{(t)}\label{ge}
\end{equation}
Equation (\ref{ge}) in the limit can be seen to lead to the
relativistic Klein-Gordon equation rather than the Schrodinger
equation \cite{bgscsfqfst}. This is an alternative justification for
our earlier result that Special Relativity emerges from the above
considerations.\\
2. We have seen that the path integral formulation is an alternative
to the Schrodinger equation, an alternative that has a resemblance
to the stochastic mechanics encountered earlier. However we should
bear in mind that these paths are merely mathematical tools for
computing the evolution of the wave functions \cite{tumulka}.
Nevertheless we should note that the path integral formulation does
not give the probability distribution on the space of all paths, so
that we cannot legitimately conclude that nature chooses one of the
several paths at random according to the probability distribution.
Unfortunately in this formulation the measures is complex and not
even rigorously defined in the limit of the continuum. Nor will the
imaginary or real paths of the measure give the actual Quantum
Mechanical picture. It would be more correct to say that the paths
are possible paths for a part of the Quantum Mechanical wave. In any
case, all this reflects via (\ref{e1b}), the unphysicality within
the minimum interval $\Delta t$.\\
On the other hand there is the well known Bohmian formulation of
Quantum Mechanics which uses the Schrodinger wave function, and the
Schrodinger equation to deduce the Hamilton-Jacobi equation exactly
as in the stochastic case. But the resemblance is superficial. This
non relativistic formulation is one in which the observer plays no
part. There is a hidden variable in the form of the position
coordinate of the particle. Thus one of the Bohmian paths represents
the actual motion of the particle, which exists separately from the
wave function. Moreover the Quantum potential $Q$ in the Bohmian
case has a non local character and no clear explanation. Furthermore
there is no clear generalization to the relativistic case. For all
these reasons though Bohm studied this approach in the 1950s, it has
not really
caught on and we will not pursue the matter further.\\
3. As mentioned discrete space time and some of their effects have
been studied from different points of view for several decades now.
It is worth mentioning here that the usual notion of time as an
operator with continuous eigen values in Quantum Theory runs into
difficulty, as was appreciated by Pauli a long time ago\cite{r125}.
This can be seen by a simple argument, and, we follow
Park\cite{r126}: Let the time operator be denoted by $\hat T$,
satisfying
$$\left[ \hat T , \hat H \right] = \imath.$$
Let $|E'>$ be an eigenfunction of $\hat H$ belonging to the
eigenvalue $E'$, and let $|E'>_\epsilon = e^{\imath \epsilon \hat
T}|E'>$. Then
\begin{equation}
\hat H | E'>_\epsilon = e^{\imath \epsilon \hat T} e^{-\imath
\epsilon \hat T} \hat H e^{\imath \epsilon \hat T}|E'> = (E' +
\epsilon )|E'>)\label{e22}
\end{equation}
Remembering that $\epsilon$ is arbitrary, (\ref{e22}) gives a
continuous energy spectrum, contrary to Quantum Theory. The
difficulty is resolved if in the
above considerations time were discrete.\\
4. It must be emphasized that in the stochastic formulation given in
this Chapter, there are no hidden variables as in the Bohm
formulation, due to the randomness or stochasticity, itself\cite{r60}.\\
5. Though we will return to some of the above considerations later,
it must be re-emphasized that in the absence of the double Weiner
process alluded to, the imaginary part of the complex velocity
potential $U$, vanishes, that is, so does $\nu$ of equation
(\ref{e3}). In this case we come back to the domain of classical
non-relativistic physics. So the origin of special relativity and
Quantum Mechanics is to be found here in this double Weiner process
within the Compton scale \cite{uof}. As pointed out in \cite{cu}
non-relativistic Quantum Mechanics is not really
compatible with Galilean or Newtonian Mechanics.\\
6. Finally, we would like to reemphasize the following point: By
neglecting terms of the order $l^2$ (the squared Compton length), we
return to point, commutative space time and can still have Quantum
Mechanics and even relativistic Quantum Mechanics and Quantum Field
Theory, though we would then have to introduce Quantum Mechanical
spin by separate arguments and consider averages over the Compton
scale anyway. But in the process, we are neglecting the Quantum of
area or Abbot and Wise's fractal dimension of the Quantum Mechanical
path. That is, we are snuffing out the fine structure implied by
Quantum Theory and are then using, as remarked earlier, a thick
brush to fudge. A quick way to see the result of Abbot and Wise is
as follows \cite{r60}. From (\ref{e3}) it follows that
$$\langle v^2 \rangle \propto (\Delta t)^{-1}$$
Now if the Hausdorf dimension \cite{r51} is $D$, we would have,
$$\Delta t = (\Delta x)^D$$
whence
$$\langle v^2 \rangle \propto (\Delta t)^{2[(\frac{1}{D})-1]}$$
A comparison yields, $D = 2$.

\end{document}